\documentclass[10pt]{article}
\usepackage[T1]{fontenc}
\usepackage[final]{epsfig}
\usepackage{amssymb}
\usepackage{multicol}
\usepackage{graphics}
\usepackage{ntimes}
\usepackage{color}

\renewcommand{~}{\,}
\newcommand{\e}{\mathtt{e}}
\newcommand{\Ham}{{\mathcal H}}
\newcommand{\dr}{\mathtt{d}r}
\newcommand{\nbar}{{\overline{n}}}
\newcommand{\BE}{\begin{equation}}
\newcommand{\EE}{\end{equation}}

\begin{document}

\title{Inter-vehicle gap statistics on signal-controlled crossroads}

\author{\textbf{Milan Krb\'alek}\\
\small Department of Mathematics\footnote{e-mail address: \textcolor{blue}{\texttt{milan.krbalek@fjfi.cvut.cz}}}\\
\small Faculty of Nuclear Sciences and Physical Engineering\\ \small Czech Technical University\\
\small Trojanova 13, 120~00 Prague, EU - Czech Republic}

\maketitle

\begin{abstract}
We investigate a microscopical structure in a chain of cars
waiting at a red signal on signal-controlled crossroads. Presented
is an one-dimensional space-continuous thermodynamical model
leading to an excellent agreement with the data measured.
Moreover, we demonstrate that an inter-vehicle spacing
distribution disclosed in relevant traffic data agrees with the
thermal-balance distribution of particles in the thermodynamical
traffic gas (discussed in \cite{Krbalek_1}) with a high inverse
temperature (corresponding to a strong traffic congestion).
Therefore, as we affirm, such a system of stationary cars can be
understood as a specific state of the traffic sample operating
inside a congested
traffic stream.\\



\end{abstract}


\section{Introduction}
\label{Introduction}

Investigation of various transport systems is recently one of the
prominent subjects of physics. Intention of relevant researches is
to describe such systems (or phenomena) quantitatively, create
their appropriate models (theo\-re\-ti\-cal or nu\-me\-ri\-cal),
and finally obtain the exact or numerical outputs comparable to
real situations. Higher aspiration of those scientific researches
can be found in finding a certain connection among the
different phenomena and revealing a possible universality.\\

Currently, one of the strongly accented fields is an investigation
of queuing systems. Under the terms of that field it has been
discussed many various topics, for example, wide-ranging spectrum
of vehicular traffic problems \cite{Review_1}, pedestrian dynamics
\cite{Review_Schreck}, escape panic \cite{Dirk_Nature},
longitudinal parking of cars on a street \cite{Abul-Magd},
\cite{Rawal}, parallel parking \cite{Paromtchik}, \cite{Arizona},
or public transport in some Latin America countries \cite{Seba},
\cite{Deift}. All those subjects are in close connection with the
Random Matrix Theory, theory of chaos, or theory of particle gases
(see the references cited above). The main goal of this paper is
to extend the set of queuing systems mentioned above by a
stationary ensemble of cars waiting at a red signal on
signal-controlled
crossroads (see also \cite{Fouladvand}).\\

Moreover, we are aiming to create an one-dimensional model of
point-like vehicles producing the same inter-vehicle gap
distributions as those detected among cars standing on
signal-controlled intersections. In the second part of this
article we demonstrate that such a model can be interpreted (on a
microscopical level) as a thermodynamical gas of dimensionless
particles exposed to a thermal bath. This analogy allows, as we
assert, to find an exact form of relevant spacing distribution
which can be consequently compared to the realistic
gap-statistics.

\section{Describing the system}
\label{Describing the system}

The traffic data analyzed in this work were measured during a few
days on a multi-lane intersection located near center of Prague.
This intersection is a constituent of an extensive network of
roads and crossroads inside the internal metropolis and is
therefore strongly saturated during the whole daytime practically.
Furthermore, the time interval between two green signals (on one
crossroad) is very short, which causes that some cars are not able
to reach the threshold of following intersection (during one green
phase) and have to wait therefore for another green light. This
fact finally leads to a substantial decrease of average velocity
of vehicles moving between crossroads, i.e one can observe the
effects usually detected in congested traffic regime (see Ref.
\cite{Review_1}). Measured were bumper-to-bumper distances $r_i$
between subsequent cars ($(i+1)$th and $i$th ones) waiting at a
red signal (in one direction only). Data file contains $5022$
digitally gauged events showing the mean inter-vehicle gap equal
approximately to 149 centimeters. We add that the clearances
were measured directly using the laser technology.\\

\begin{figure}
\label{fig:2} \centering
\scalebox{.4}{\includegraphics{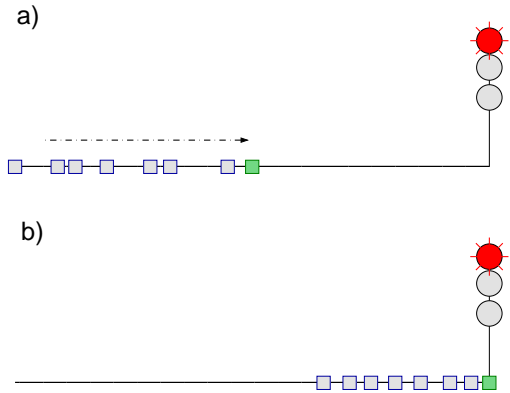}}
\parbox{7cm}{\caption{\footnotesize Graphic representation of the model. The upper subplot depicts the initial state
of the numerical scheme described in text whereas the bottom
subplot demonstrates the final stationary state of the traffic
sample, i.e the state when the cars are waiting for green signal.
We note that the squares represent the model particles with the
leading car being picked out.}}
\end{figure}

More detailed statistical analysis uncovers that a probability
density $p(r)$ for distance $r$ between neighboring cars shows a
similar behavior as that investigated between the eigenvalues of
random matrices (see \cite{Mehta}), zeros of Riemann zeta function
(see \cite{Bogomolny}), or vehicles moving inside the traffic
stream on the freeways (see \cite{Krbalek_1}). Such a behavior
(follow Fig. 2) demonstrates the presence of repulsive
interactions among the elements in question. As well known, a
spacing distribution of non-interacting elements shows a different
distribution, in concrete: \emph{Poisson probability density}
$$p(r)=\exp[-r]\hspace{1cm}(r\geq0).$$
Since the traffic interaction (in local sense, of course) is
usually quantified as power-law repulsion among the successive
vehicles (see Ref. \cite{Krbalek_1} and \cite{Dirk}) let us
suppose that a potential energy of the ensemble investigated reads
as

\BE U(r_1,r_2,\ldots,r_N)=\sum_{i=1}^N
r_i^{-1}.\label{potential_energy}\EE
Herein we assume that the stationary traffic state analyzed in
this paper (i.e. the queue of waiting cars) is determined by the
preceding process -- traffic flow towards the intersection.
Evidently, moving in traffic sample the driver is interacting with
other cars and optimizing his/her motion to reach the threshold of
the crossroad as soon as possible and, at the same time, avoid a
crash with the preceding vehicle. Such a behavior corresponds to
the thermodynamic effects governing the ensemble into a local
thermal equilibrium (see \cite{Krbalek_1} for details).

\section{Modified Metropolis algorithm}
\label{Modified Metropolis algorithm}

\begin{figure}
\label{fig:5} \centering
\scalebox{.4}{\includegraphics{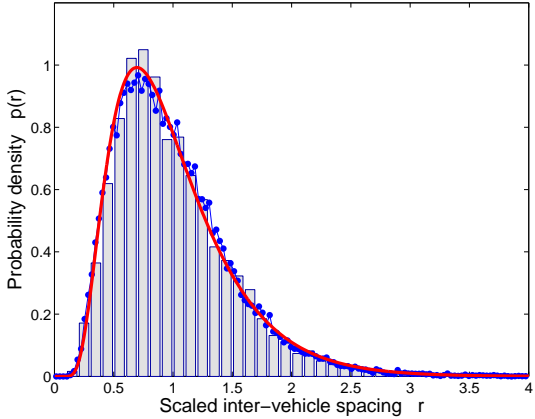}}
\parbox{7cm}{\caption{\footnotesize Inter-vehicle gap statistics $p(r).$ Bars represent the probability
density for bumper-to-bumper distance among the cars waiting at a
red signal on intersections (measured in Prague). Data were
re-scaled so that the mean spacing is equal to one. Points display
the optimized result of the numerical scheme (Metropolis
algorithm) for $N=100$ and $\beta_{model}=1.45.$ Finally, the
curve displays the distribution (\ref{gap-distri}) for the fitted
value $\beta_{fit}\approx1.2488$ (obtained by the number variance
test.)}}
\end{figure}

Accepting the above-mentioned assumptions on thermodynamical
aspects of the issue we formulate the following one-dimensional
traffic model based on principles of statistical physics. Consider
$N+1$ point-like particles (cars) located randomly (or
equidistantly if advantageous) on a line (or on a circle) so that
the mean gap among them is one, i.e.
\BE \sum_{i=1}^N r_i=N,\label{sumri}\EE
where $r_i$ represents the gap between $(i+1)$th and $i$th
particles. Thus, the ordered positions $x_1>x_2>\ldots>x_{N+1}$
constitute the initial state for our simulation (see Fig. 1). The
particles move along the line (or along the circle) accepting the
undermentioned rules until the leading car reaches a fixed point
(the threshold of new crossroads). In accord to a realistic
situation the overtaking cars are not permitted, i.e. the
particles can not change their order. Let $\beta_{model}\geq 0$
denote the inverse temperature specifying the measure of chaos
inside the ensemble simulated. We assume $\beta_{model}$ to be the
only significant parameter of the model. The car positions are
repeatedly updated (we use 20000 steps in our version) according
to the following rules:

\begin{enumerate}
\item Calculated is the potential energy $U_0$ (using formula
(\ref{potential_energy})) for the actual set of locations
$\{x_1,x_2,\ldots,x_{N+1}\}.$
\item We pick an index $j\in\{1,2,\ldots,N+1\}$ at random.
\item We draw a random number $\delta$ equally distributed in
the interval $(0,1).$
\item We compute an anticipated position $x_j'=x_j+\delta$ of
$j$th element. Because of singularity in the potential energy
(\ref{potential_energy}) the model particles can not change their
order. Therefore we accept $x_j'$ only if $x_j'<x_{j-1}.$
\item We calculate a value of potential energy $U'$ determined for configuration
$\{x_1,x_2,\ldots,x_{j-1},x_j',x_{j+1},\ldots, x_{N+1}\}.$
\item If $U' \leq U_0$ the $j$th particle position take on a new
value $x_j'.$ If $U' > U_0$ then the Boltzmann factor
$$w=\exp\left[-\beta_{model}\Delta U\right],$$ where $\Delta U=U'-U_0,$ should be compared with
a random number $r$ equally distributed in $(0,1).$ Provided that
the inequality $w > r$ is fulfilled the $j$th particle position
takes on the new value $x_j'$ too. Otherwise, the original
configuration $\{x_1,x_2,\ldots,x_{N+1}\}$ remains unchanged.
\end{enumerate}

The sketched procedure represents a modified Metropolis algorithm
originally developed for chemistry purposes (in Ref.
\cite{Metropolis}). This algorithm belongs to the category of
Monte Carlo simulations (see Ref. \cite{Landau}) which are
recently used for numerical modelling of statistical systems (as
demonstrated in Ref. \cite{Scharf}, for example). The elaborated
scheme of Metropolis ensures a relaxation of ensemble into a
thermal-balance state when the energy fluctuates around a constant
value being independent of initial configuration of particles (see
Fig. 3). After reaching the thermal equilibrium (i.e. after
approximately 5000 updates of configuration (Monte Carlo steps),
as visible in Fig. 3) the ensemble lingers in this state until the
simulation is interrupted. Then, as observed, corresponding
probability density for inter-particle gaps depends
on the inverse temperature $\beta_{model}$ only.\\

Our aim is to find the optimal value of inverse temperature
$\beta_{model}$ so that the gap distribution $p(r)$ corresponds to
that measured among the cars on crossroads. Using a
$\chi^2-$method (i.e. minimizing the sum of squares-deviations
between two distributions in question) one can find that optimal
value $\beta_{model}$ is approximately 1.45. Concretely, for fixed
value $\beta_{model}$ the distribution $p(r)$ is obtained. Then
the $\chi^2-$test between empirical data and $p(r)$ could be
evaluated. The optimal value of $\beta_{model}$ is the one for
which the corresponding sum of squares-deviations is minimal. To
conclude, for value $\beta_{model}=1.45$ both processes (traffic
and Metropolis procedure) generate practically the same gap
distributions (see Fig. 2). Thus, the introduced procedure could
represent a realistic model for behavior of the cars in the
vicinity of the chosen intersection.

\begin{figure}
\label{fig:4} \centering
\scalebox{.4}{\includegraphics{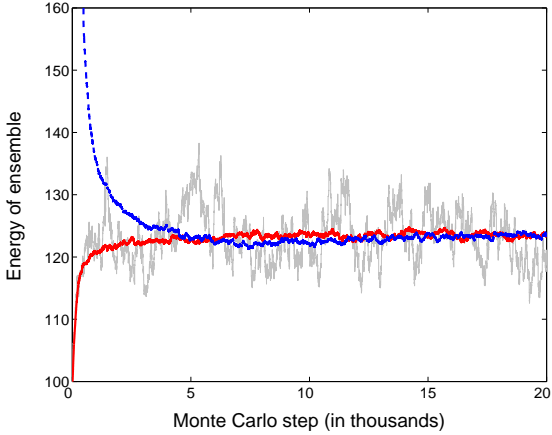}}
\parbox{7cm}{\caption{\footnotesize Relaxation of the system into the thermal
equilibrium. Dashed and continuous lines (see the upper-left or
lower-left corners, respectively) display the energy value
(\ref{potential_energy}) for $N=100$ and $\beta_{model}=1.45$
during the run of Metropolis procedure (having 20000 steps) for
random (or equidistant) initial locations of elements,
respectively. Plotted is the average value (calculated for 100
repeated realizations of Metropolis). Grey curve represents the
energy value (\ref{potential_energy}) for one realization of
Metropolis (when initial particle positions were chosen
equidistantly).}}
\end{figure}

\section{Terminal state of thermodynamical traffic gas}
\label{Terminal state of thermodynamical traffic gas}

As explored in articles \cite{Dirk}, \cite{Krbalek_1}, and
\cite{Krbalek_2}, the traffic flow can be understood (on a
microscopical level) as a thermodynamical gas of interacting cars
exposed to a heat bath of inverse temperature $\beta.$ Besides,
the latter has an immediate relation to the traffic density. If
accepting such a approach we describe the traffic ensemble (on the
move) as a circular gas of point-like particles whose hamiltonian
reads as
$$\Ham=\sum_i (v_i-\overline{v})^2+\sum_i
r_i^{-1},$$
where $v_i$ and $r_i$ represent a $i$th car velocity and gap to
the previous car, respectively. Quantity $\overline{v}$ denotes
the desired velocity of the ensemble. Then (see the exact
calculation in \cite{Krbalek_1}) the derived probability density
$p_\beta(r)$ for a gap $r$ among the successive vehicles is
\BE p_\beta(r)=A\exp\left[\beta r^{-1}-Br\right],
\label{gap-distri} \EE
where the constants $A$ and $B$ are calculated via two
normalization equations
$$\int_0^\infty p_\beta(r)~\dr=\int_0^\infty r~p_\beta(r)~\dr=1.$$
According to Ref. \cite{Krbalek_1} the following relations hold
true:
$$B\approx \beta  + \frac{3-\e^{-\sqrt{\beta}}}{2},$$

$$A \approx
\frac{\sqrt{2\beta+3-\e^{-\sqrt{\beta}}}}{\sqrt{8\beta}~\mathcal{K}_1\left(\sqrt{4\beta^2+6\beta-2\beta
\e^{-\sqrt{\beta}}}\right)}.$$
Herein ${\mathcal{K}}_1$ stands for a Mac-Donald's function
(modified Bessel's function of the second kind) of the first order.\\

Since the situation investigated in this article is without any
doubt the result of a preceding traffic flow (see
\cite{Fouladvand}) it is meaningful to expect that the clearance
distribution among the cars waiting at the red-light-signal will
be of the form (\ref{gap-distri}). Indeed, as confirmed by an
appropriate statistical analysis of the collected data (discussed
later) the measured gap statistics (see Fig. 2) corresponds to the
probability density (\ref{gap-distri}) if the inverse temperature
$\beta$ of the thermodynamical model is
\BE \beta_{fit}\approx1.2488. \label{fitka} \EE
We denote that this value has been determined by a more
sophisticated method presented in next section. In addition, a
positive comparison between the corresponding gap distributions
supports the hypothesis that traffic stream can be \emph{locally}
understood as a stochastic gas whose elements are repulsed by the
forwardly-directed nearest-neighbor power-law potential depending
on a reciprocal distance between successive gas-elements. This
correspondence, however, does not mean that traffic is a
thermodynamical system, of course.

\section{Testing the statistical variance of data}
\label{Testing the statistical variance of data}

If trying to find a more robust argumentation for an assertion on
statistical similarities between the process investigated and the
traffic model we can apply some of the techniques originally
developed for purposes of the Random Matrix Theory (see the book
\cite{Mehta}). Usual way how to quantify the behavior of variances
among the statistical data is in applying so-called
\emph{number-variance test.} Such a test is defined as follows.\\

Consider $N$ spacings $r_1,r_2,\ldots,r_N$ between the successive
vehicles (or particles of model) and suppose that the mean
distance taken over the complete ensemble is re-scaled to one,
i.e.
$$\sum_{i=1}^N r_i=N.$$
Dividing the interval $[0,N]$ into subintervals $[(k-1)L,kL]$ of a
length $L$ and denoting by $n_k(L)$ the number of cars in the
$k$th subinterval, the average value $\nbar(L)$ taken over all
possible subintervals is
$$\nbar(L)=\frac{1}{\lfloor N/L \rfloor} \sum_{k=1}^{\lfloor N/L
\rfloor} n_k(L)=L,$$
where the integer part $\lfloor N/L \rfloor$ stands for the number
of all subintervals included in the interval $[0,N].$ Number
variance $\Delta_n(L)$ is then defined as
$$\Delta_n(L)=\frac{1}{\lfloor N/L \rfloor} \sum_{k=1}^{\lfloor N/L
\rfloor} \left(n_k(L)-L\right)^2$$
and represents (in a traffic instance) the statistical variance in
the number of vehicles operating at the same time inside a fixed
part of the road of a length $L.$\\

\begin{figure}
\label{fig:6} \centering
\scalebox{.4}{\includegraphics{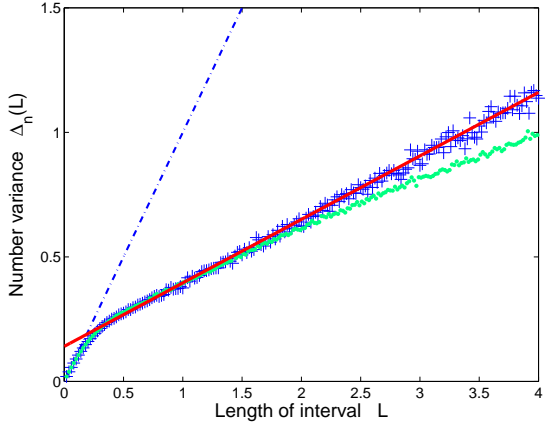}}
\parbox{7cm}{\caption{\footnotesize Results of the number-variance test.
Plus signs stand for the value $\Delta_n(L)$ calculated for the
collected traffic data, whereas points represent the same quantity
for particles of Metropolis model (where $\beta_{model}=1.45$).
Dash-dotted line visualizes function $\Delta_n(L)=L$ representing
the number variance of independent events. The solid curve
displays the function (\ref{vztah_pro_NV}) calculated for the
optimal value $\beta_{fit}\approx1.2488$ obtained by the
$\chi^2-$fit to the traffic data.}}
\end{figure}
As well known from Random Matrix Theory the number variance can be
explicitly derived from the relevant spacing density $p_\beta(r).$
The significant advantage is remarkable sensitivity of the number
variance $\Delta_n(L)$ to any change in the probability density
$p_\beta(r)$ -- i.e. to any change in the potential
$U(r_1,r_2,\ldots,r_n)$ also. Whereas the number variance of
independent events (or non-interacting elements) is the identity
$\Delta_n(L)=L,$ for a thermodynamical traffic gas with non-zero
inverse temperature $\beta$ there has been numerically calculated
(in Ref. \cite{Krbalek_2}) a different behavior, concretely: a
linear dependence
\BE \Delta_n(L)\approx\chi L+\gamma \label{vztah_pro_NV} \EE
with a slope
$$\chi \approx \frac{1}{2.4360~\beta^{0.8207}+1}\leq 1$$
and a shift
$$\gamma \approx \frac{\beta}{5.1926~\beta+2.3929}\geq 0.$$
As understandable now, the comparison between the number variance
of the collected data and the function (\ref{vztah_pro_NV}) can be
then used (together with the comparison of the relevant gap
distributions) as a robust fitting procedure which is capable of
revealing more detailed nuances among the distributions compared.
If applied to our topic, such a procedure generates the optimal
value (\ref{fitka}) for which the exactly determined number
variance (\ref{vztah_pro_NV}) corresponds to the measured data
(see Fig. 4). Note that both of curves $\Delta_n(L)$ are rapidly
deflected from the line visualizing the number variance of
non-interacting particles. It implies the presence of a strong
repulsion among the vehicles. However, a small deviation is
detected for larger $L$ between the traffic data (plus signs in
Fig. 4) and Metropolis data (points in the same figure). Such a
discrepancy can be explained by the simple fact that the
respective temperatures (i.e. $\beta_{model}$ and $\beta_{fit}$)
differ each from other.

\section{Summary and discussion}
\label{Summary and discussion}

Investigated was the traffic ensemble of vehicles waiting at a
red-light-signal on signal-controlled crossroads. We have
introduced the thermal space-continuous time-discrete traffic
model of repulsing point-like elements based on the Metropolis
algorithm. By the suitable choice of the inverse temperature
parameter there were obtained the same statistical distributions
as those produced by the real traffic process. Above that, we show
that the investigated state of the realistic traffic sample can be
predicted with the help of the thermal-equilibrium state for local
thermodynamical gas whose point-like particles are repulsed by the
short-range power-law potential (\ref{potential_energy}). As
demonstrated above, for the fitted value $\beta\approx1.25$ of
reciprocal temperature the corresponding spacing distributions are
practically the same. The correspondence between the traffic
samples and presented theory is, moreover, supported by the robust
test of number variance which reveals
\begin{enumerate}
\item the thermal feature of the topic -- on microscopic scale,
\item the presence of strong interactions among the cars,
\item a deep connection between the stationary state of waiting cars and the
preceding move of sample towards the intersection
threshold.\end{enumerate}
To conclude, we assert that the configuration of vehicles waiting
at a red-light-signal on signal-controlled crossroads is a product
of local thermodynamics-like processes acting among the cars. All
the accessible statistical analyses strongly support this fact.
Therefore, the observed phenomenon can be understood as a
traffic in especial super-congested state.\\

\emph{Acknowledgement} This work was supported by the Ministry of
Education, Youth and Sports of the Czech Republic within the
project MSM 6840770039. We would like to thank the students of
FNSPE (attending the seminar 01SMAB on calculus, academic year
2006/2007) who collected the traffic data analyzed in this
article.

\end{document}